\documentclass[aps,prc,groupedaddress,showpacs,eqsecnum]{revtex4}
\usepackage{graphicx}
\usepackage{dcolumn}
\usepackage{bm}

\newcommand{\raa}{($\alpha$,$\alpha$)}

\newcommand{\rga}{($\gamma$,$\alpha$)}
\newcommand{\rag}{($\alpha$,$\gamma$)}

\newcommand{\oaa}{$^{16}$O\raa $^{16}$O}
\newcommand{\oag}{$^{16}$O\rag $^{20}$Ne}
\newcommand{\nega}{$^{20}$Ne\rga $^{16}$O}

\newcommand{\sfac}{S-factor}

\begin{document}

\title{
Low-Energy Direct Capture in the
$^{16}$O($\bm\alpha$,$\bm\gamma$)$^{20}$Ne Reaction
}

\author{P.\ Mohr}
\email[E-mail: ]{WidmaierMohr@compuserve.de}
\affiliation{
Diakoniekrankenhaus Schw\"abisch Hall, D-74523 Schw\"abisch Hall,
Germany
}

\date{\today}

\begin{abstract}
The cross section of the \oag\ capture reaction is analyzed at low
energies where the direct capture mechanism is dominant. For
temperatures below $T_9 = 0.2$ the resulting astrophysical reaction
rate is about a factor of two higher than in a recent compilation
whereas the energy dependence of the astrophysical \sfac\ and the
branching ratios to the $^{20}$Ne bound states are very similar to
previous calculations. The validity of the widely used detailed
balance theorem for the inverse \nega\ photodisintegration rate is
confirmed for the special case of high-lying first excited states.
\end{abstract}


\maketitle

\section{\label{sec:intro}Introduction}
The astrophysical reaction rate of the \oag\ reaction is dominated by
a series of resonances in a broad range of high temperatures above
$0.2 \times 10^9$\,K ($T_9 \ge 0.2$). However, at low temperatures
($T_9 < 0.2$) the direct capture mechanism becomes dominant
\cite{NACRE}. 

The \oag\ reaction is an important reaction during helium burning in
stars at typical temperatures around $T_9 = 0.2$ which is the
temperature range where both direct capture and resonances are
important. The \oag\ reaction blocks the reaction chain
3\,$\alpha$ $\rightarrow$ $^{12}$C\rag \oag\ because there are no
resonances in the so-called Gamow window which is located at $E_0 \approx
1.22 \left( Z_1^2 Z_2^2 A_{\rm{red}} T_6^2 \right)^{1/3}\,{\rm{keV}}$
with a width of $\Delta = 0.749 \left( Z_1^2 Z_2^2 A_{\rm{red}} T_6^5
\right)^{1/6}\,{\rm{keV}}$; $Z_1$ ($A_1$) and $Z_2$ ($A_2$) are the
charge (mass) numbers of projectile and target, and $A_{\rm{red}} =
A_1 A_2/(A_1+A_2)$ is the reduced mass number. At $T_9 = 0.2$ one
finds $E_0 = 390$\,keV and $\Delta = 190$\,keV. In several previous
papers the \sfac\ at $E_0 = 300$\,keV is reported. The $2^-$ state at
$E_x = 4967$\,keV in $^{20}$Ne does not appear as resonance at $E =
237$\,keV in the \oag\ reaction because of its unnatural parity.

The inverse \nega\ reaction plays an important role in neon burning at
significantly higher temperatures around $T_9 = 1 - 2$
\cite{Thi85}. The reaction rate of photodisintegration reactions is
usually derived from the capture rate using the detailed balance
theorem (see e.g.\ \cite{NACRE}). The validity of this theorem for the
case of the \oag\ and \nega\ reactions with high-lying first excited
states is analyzed in the last section \ref{sec:inv}.

Direct capture in the \oag\ capture reaction has been analyzed
theoretically by \cite{Du94,Ba88b,De86,Ba85,De83,La83}. Experimentally
a non-resonant \sfac\ of $S = 0.26 \pm 0.07$\,MeV\,b for the ground
state transition has been reported at energies around $E = 2$\,MeV
\cite{Ha87}, and a total \sfac\ of $S(300\,{\rm{keV}}) < 0.7$\,MeV\,b
was deduced, but this result has been questioned by
\cite{Ba88b}. Later, an upper limit of $S < 3.37$\,MeV\,b was reported
\cite{Kn94} which was derived by several data points between 1.4\,MeV
and 2.4\,MeV; no extrapolation to $S(300\,{\rm{keV}})$ is given in
\cite{Kn94} because of the contradicting energy dependencies of the
\sfac\ in \cite{La83,Ba88b}. The theoretical predictions vary between
0.7\,MeV\,b and 2.5\,MeV\,b at 300\,keV
\cite{Du94,Ba88b,De86,Ba85,De83,La83}.

This work gives a new prediction for the direct capture cross section
of the \oag\ reaction at low energies. The calculation is based on a
systematic folding potential which is able to reproduce simultaneously
scattering data for $^{16}$O\raa $^{16}$O and bound state properties
of the system $^{20}$Ne = $^{16}$O $\otimes$ $\alpha$
\cite{Abe93}. The direct capture model is presented in
Sect.~\ref{sec:model}. Relevant bound state properties are listed in
Sect.~\ref{sec:bound}. The direct capture cross section is shown in
Sect.~\ref{sec:results}, and the results are discussed in
Sect.~\ref{sec:disc}. After a brief analysis of the inverse
\nega\ photodisintegration reaction in Sect.~\ref{sec:inv},
conclusions are drawn in Sect.~\ref{sec:summconc}.

\section{\label{sec:model}Direct Capture, $B(E/M{\cal{L}})$ values,
  and Folding Potentials} 

\subsection{\label{sec:DC}Direct Capture}
The direct capture (DC) cross section for $E1$, $E2$, and $M1$
transitions is given in detail in Refs.~\cite{Kim87,Mohr93}. Here only
the essential properties of the model are repeated. The DC cross
section is proportional to the square of the overlap integral of the
scattering wave function $\chi_l$, the electromagnetic transition
operator ${\cal{O}}^{E/M{\cal{L}}}$, and the bound state wave function
$u_{NL}$:
\begin{equation}
\sigma_{DC}(E) \sim \left| \int dr \, u_{NL}(r) \,
{\cal{O}}^{E/M{\cal{L}}} \, \chi_l(r,E) \right|^2
\label{eq:DC}
\end{equation}
The radial parts of the dominating E1, E2, and M1 electromagnetic
transition operators are given by:
\begin{eqnarray}
{\cal{O}}^{E1} & = & \frac{3}{\rho^3} \left[ (\rho^2-2)\sin{\rho} + 2 \rho
\cos{\rho} \right] r \approx r \label{eq:opE1} \\
{\cal{O}}^{E2} & = & \frac{15}{\rho^5} \left[ (5\rho^2-12) \sin{\rho}
  + (12 - \rho^2) \rho \cos{\rho} \right] r^2 \approx r^2 \label{eq:opE2}\\
{\cal{O}}^{M1} & = & \frac{1}{2\rho} \left[ \sin{\rho} + \rho
\cos{\rho} \right] \approx 1 \label{eq:opM1}
\end{eqnarray}
with the so-called long wavelength approximation for $\rho = k_\gamma
r \ll 1$. From Eq.~(\ref{eq:opM1}) and from the orthogonality of the
wave functions it is obvious that the $M1$ contributions will be very
small in the DC model (see Sect.~\ref{sec:results}). The exact form of
the multipole operators in Eqs.~(\ref{eq:opE1}), (\ref{eq:opE2}), and
(\ref{eq:opM1}) was used in the following calculations. 

The coefficient of proportionality in Eq.~(\ref{eq:DC}) for $E1$
transitions contains a factor
\begin{equation}
\tilde{C}(E1) = \left[ C(E1) \right]^2 \sim \left[ \frac{A_1 A_2}{A_1
    + A_2} \left( \frac{Z_1}{A_1} - \frac{Z_2}{A_2} \right) \right]^2
\label{eq:isoE1}
\end{equation}
which describes the isospin suppression of $E1$ transitions in $N = Z$
nuclei between $T=0$ states. If one uses integer mass numbers, the
$E1$ cross section exactly vanishes; using precise masses, one
calculates very small $E1$ cross sections. In reality, small $T=1$ isospin
admixtures in the scattering and bound state wave functions may lead
to a significant enhancement of the capture cross section compared to
the DC model calculation \cite{De86}.

\subsection{\label{sec:BE}$B(E/M{\cal{L}})$ values}
Whereas Eq.~(\ref{eq:DC}) determines the overlap integral at an
arbitrary energy $E$ of the scattering wave function, the following
Eq.~(\ref{eq:BE}) gives a similar overlap integral between
bound state wave functions at their defined energies $E_{x,i}$ and
$E_{x,f}$ for transitions $L_i \rightarrow L_f$ \cite{Buck77,Hoy94}: 
\begin{equation}
B(E{\cal{L}},L_i \rightarrow L_f) =
	\frac{e^2\beta_{\cal{L}}^{2} (2L_f+1) (2{\cal{L}}+1)}{4\pi (2L_i+1)} 
	\cdot
	\langle L_f 0 {\cal{L}} 0 | L_i 0 \rangle ^2
	 \, \, \, \times
	\left[ \int_{0}^{\infty}
	u_{N_fL_f}(r) r^{\cal{L}} u_{N_iL_i}(r) \; dr \right] ^{2}
\label{eq:BE}
\end{equation}
with
\begin{equation}
\beta_{\cal{L}} = 
	\frac{A_2^{\cal{L}} Z_1 + (-1)^{\cal{L}} A_1^{\cal{L}} Z_2}
		{(A_1+A_2)^{\cal{L}}}
\label{eq:betaL}
\end{equation}
The above formalism is also valid for quasi-bound states with $E > 0$
which appear as resonances in the capture reaction.

Obviously, one finds the same suppression of $E1$ transitions between
$T=0$ states in $N=Z$ nuclei because the suppression factors in
Eq.~(\ref{eq:isoE1}) and (\ref{eq:betaL}) are identical:
$C(E1) = \beta_1$. This identity
gives the chance to renormalize the DC calculation by the ratio of
experimental and calculated $B(E{\cal{L}})$ values
\begin{equation}
r_{BE} = \frac{B_{\rm{exp}}(E{\cal{L}})}{B_{\rm{calc}}(E{\cal{L}})}
\label{eq:ratio}
\end{equation}
leading to a theoretically predicted capture cross section
$\sigma_{\rm{th}}$: 
\begin{equation}
\sigma_{\rm{th}}(E) = r_{BE} \times \sigma_{DC}(E)
\label{eq:sigma}
\end{equation}

One expects ratios $r_{BE} \approx 1$ for $E2$ transitions and
$r_{BE} \gg 1$ for $E1$ transitions. Because $M1$ transitions play
only a very minor role (see Sect.~\ref{sec:results}), $r_{BM} = 1$ was
used in all cases. The expectations for the $r_{BE}$ values are
confirmed by the calculations in Sect.~\ref{sec:bound}.

\subsection{\label{sec:fold}Folding Potentials}
The essential ingredients for the calculation of the capture cross
section and the $B(E{\cal{L}})$ values are the potentials which are
required for the calculation of the wave functions in
Eqs.~(\ref{eq:DC}) and (\ref{eq:BE}). Based on the double-folding
approach, a potential has been derived in \cite{Abe93} which is
capable of describing simultaneously elastic $^{16}$O\raa $^{16}$O
scattering at energies above and below the Coulomb barrier and bound
state properties of the system $^{20}$Ne = $^{16}$O $\otimes$
$\alpha$.

The potential $V(r)$ is given by
\begin{equation}
V(r) = V_C(r) + \lambda V_F(r)
\label{eq:pot}
\end{equation}
with the Coulomb potential $V_C(r)$ of a homogeneously charged sphere
with radius $R_C$ and the folding potential $V_F(r)$ which is scaled
by a strength parameter $\lambda$ of the order of $\lambda \approx 1.2
- 1.3$.
\begin{equation}
V_F(r) = 
	\int \int \rho_1(r_1) \, \rho_2(r_2) \, v_{\rm{eff}}(E) \;
	d^3r_1 \; d^3r_2 
\label{eq:dfold}
\end{equation}
$\rho_{1,2}$ are the densities of the $\alpha$ particle and $^{16}$O
which are derived from electron scattering data \cite{Vri87}, and
$v_{\rm{eff}}$ is an effective interaction \cite{Abe93,Kob84}. The
Coulomb radius $R_C$ has been chosen identical to the root-mean-square
radius of the folding potential ($r_{\rm{rms}} = 3.603$\,fm).

\section{\label{sec:bound}Bound and quasi-bound state properties of
  $^{20}$N{\lowercase{e}} = $^{16}$O $\otimes$ $\alpha$} 
The wave function $u_{NL}(r)$ which describes the relative motion of
the $\alpha$-nucleus system is characterized by the node number $N$
and the orbital angular momentum $L$. These values are related to the
corresponding quantum numbers $n_i$ and $l_i$ of the four nucleons
forming the $\alpha$ cluster in the $sd$-shell:
\begin{equation}
Q = 2N + L = \sum_{i=1}^{4} (2n_i + l_i) = \sum_{i=1}^{4} q_i
\label{eq:Q}
\end{equation}
One expects $Q = 8$ for the ground state band and $Q=9$ for the first
negative parity band. The following procedure has been used for the
calculation of the wave functions and their properties. First, the
strength parameter $\lambda$ of the folding potential was adjusted to
reproduce the energy of the state under consideration. Second, for
resonances (which are quasi-bound states with $E > 0$) the theoretical
width $\Gamma_\alpha^{\rm{th}}$ was derived from scattering phase
shifts $\delta_L$ by 
\begin{equation}
\Gamma_\alpha^{\rm{th}} = \frac{2}{(d\delta_{L}/dE) |_{E=E_{\rm{res}},
    \delta_{L} = \pi/2}} 
\label{eq:width}
\end{equation}
The results are listed in Table \ref{tab:bound}.
\begin{table}
  \caption{\label{tab:bound} 
    Properties of the $Q=8$ and $Q=9$ states in
    $^{20}$Ne. Experimental values have been taken from
    \cite{Til98}. There is no clear assignment of $7^-$ and $9^-$
    states to the $0^-$ band in \cite{Til98}; the lowest tentative
    assignment was used. Minor numerical
    differences to Ref.~\cite{Abe93} are 
    a consequence of a different Coulomb radius $R_C$.
}
\begin{center}
\begin{tabular}{rrcccccr@{$\pm$}lr}
\multicolumn{1}{c}{$E_x$} & \multicolumn{1}{c}{$E$} & $J^\pi$ & $Q$ &
  $N$ & $L$ & $\lambda$ & 
  \multicolumn{2}{c}{$\Gamma_\alpha^{\rm{exp}}$} &
  \multicolumn{1}{c}{$\Gamma_\alpha^{\rm{th}}$} \\ 
\multicolumn{1}{c}{(keV)} & \multicolumn{1}{c}{(keV)} & & & & & &
\multicolumn{2}{c}{(keV)} & \multicolumn{1}{c}{(keV)} \\ 
\hline
0    & $-4730$ & $0^+$ & 8 & 4 & 0 & 1.2413 & \multicolumn{2}{c}{$-$} & $-$ \\
1634 & $-3096$ & $2^+$ & 8 & 3 & 2 & 1.2239 & \multicolumn{2}{c}{$-$} & $-$ \\
4248 & $ -482$ & $4^+$ & 8 & 2 & 4 & 1.2176 & \multicolumn{2}{c}{$-$} & $-$ \\
8778 & $+4048$ & $6^+$ & 8 & 1 & 6 & 1.1996 & 0.11 & 0.02 & 0.73 \\
11951& $+7221$ & $8^+$ & 8 & 0 & 8 & 1.2622 & 35  & 10\,eV & 508\,eV \\
5788 & $+1058$ & $1^-$ & 9 & 4 & 1 & 1.2663 & 28  & 3\,eV  & 40\,eV \\
7156 & $+2426$ & $3^-$ & 9 & 3 & 3 & 1.2774 & 8.2 & 0.3 & 11.9 \\
10262& $+5532$ & $5^-$ & 9 & 2 & 5 & 1.2745 & 145 & 40  & 186 \\
13692& $+8962$ & $7^-$ & 9 & 1 & 7 & 1.3251 & 158 &
18\footnote[1]{derived from $\Gamma_\alpha = 310 \pm 30$\,keV and
  $\Gamma_{\alpha_0}/\Gamma_\alpha = 0.51 \pm 0.03$}  & 147 \\
17430& $+12700$& $9^-$ & 9 & 0 & 9 & 1.4138 & 52.8 &
7.0\footnote[2]{derived from $\Gamma_\alpha = 220 \pm 25$\,keV and 
  $\Gamma_{\alpha_0}/\Gamma_\alpha = 0.24 \pm 0.01$}  & 23.6 \\
\hline
\end{tabular}
\end{center}
\end{table}

One finds that the strength parameter $\lambda$ is constant within
about $\pm 2\,\%$ for the ground state band with positive parity and
within $\pm 5\,\%$ for the negative parity band (or $\pm 2\,\%$ if one
neglects the $9^-$ state with its tentative assignment to this
band). The average values $\lambda_{\rm{even}}$ for the positive
parity band are slightly smaller than the average $\lambda_{\rm{odd}}$
for the negative parity band. Additionally, a remarkable agreement is
found between the calculated widths $\Gamma_\alpha^{\rm{th}}$ and the
experimental widths $\Gamma_\alpha^{\rm{exp}}$ especially for the
negative parity band; for the ground state band one finds an
overestimation of the order of a factor of 6.6 (14.5) for the $6^+$
($8^+$) state.

It is interesting to note that the potential for the $L=0$ partial
wave roughly reproduces also the properties of the $0^+_4$ state in
$^{20}$Ne at $E_x \approx 8700$\,keV. Using exactly the same potential
as for the ground state ($\lambda = 1.2413$) one obtains a broad
resonance with $Q=10$ at $E_x^{\rm{th}} \approx 8080$\,keV with a
width of several MeV. This compares well with the experimental values
of the excitation energy of $E_x^{\rm{exp}} \approx 8700$\,keV and of
the experimental width $\Gamma_\alpha^{\rm{exp}}> 800$\,keV. However,
it is difficult to define the excitation energy of this broad
resonance from the potential model calculation because the calculated
phase shift does not reach $\delta_0 = \pi/2$. Instead of
Eq.~(\ref{eq:width}) a revised definition for broad resonances was
used here: 
\begin{equation}
\Gamma_\alpha^{\rm{th}} = \frac{2}{(d\delta_{L}/dE) |_{E=E_{\rm{res}},
    (d\delta_{L}/dE) = {\rm{max.}}}} 
\label{eq:widthbroad}
\end{equation}

In the same way, the properties of the broad $2^+$ state in the
$0^+_4$ band are reproduced: using $\lambda = 1.2239$, one finds
$E_x^{\rm{th}} = 9130$\,keV and $\Gamma_\alpha^{\rm{th}} \approx
5$\,MeV which compares to the experimental values $E_x^{\rm{exp}} =
9000$\,keV and $\Gamma_\alpha^{\rm{exp}} \approx 800$\,keV.

Using the potential strength parameters $\lambda$ as given in Table
\ref{tab:bound}, one may calculate $B(E{\cal{L}})$ values using
Eq.~(\ref{eq:BE}) and the formalism in Sect.~\ref{sec:BE} without free
parameters. The results are listed in Table \ref{tab:BE} which
includes also the ratio $r_{BE}$ between the experimental and
theoretical values as defined in Eq.~(\ref{eq:ratio}).
\begin{table}
  \caption{\label{tab:BE} 
    $B(E{\cal{L}})$ values for electromagnetic transitions in
    $^{20}$Ne. Experimental values have been taken from
    \cite{Til98}. 
}
\begin{center}
\begin{tabular}{crcrcccr}
$J^\pi_i$ & \multicolumn{1}{c}{$E_{x,i}$} & $J^\pi_f$ &
  \multicolumn{1}{c}{$E_{x,f}$} & $\cal{L}$ &
  $B_{\rm{exp}}(E{\cal{L}})$ & $B_{\rm{th}}(E{\cal{L}})$ & 
  \multicolumn{1}{c}{$r_{BE}$} \\ 
& \multicolumn{1}{c}{(keV)} & & \multicolumn{1}{c}{(keV)} & & (W.u.) &
  (W.u.) & \\
\hline
$2^+$ &  1634 & $0^+$ &     0 & 2 & 20.3(13) & 15.6 & 1.30~~ \\
$4^+$ &  4248 & $2^+$ &  1634 & 2 & 21.9(22) & 21.0 & 1.05~~ \\
$6^+$ &  8778 & $4^+$ &  4248 & 2 & 19.7(30) & 19.7 & 1.00~~ \\
$8^+$ & 11951 & $6^+$ &  8778 & 2 &  9.0(13) & ~9.5 & 0.95~~ \\
$3^-$ &  7156 & $1^-$ &  5788 & 2 & 50.1(78) & 41.6 & 1.20~~ \\
$1^-$ &  5788 & $0^+$ &     0 & 1 & $8.3(31)\times 10^{-6}$ &
$3.9\times 10^{-6}$ & 2.13~~ \\
$1^-$ &  5788 & $2^+$ &  1634 & 1 & $1.1(2)\times 10^{-4}$ &
$7.3\times 10^{-6}$ & 15.0~~~ \\
$3^-$ &  7156 & $2^+$ &  1634 & 1 & $-$ &
$5.6\times 10^{-6}$ & 2.13\footnote[1]{$r_{BE}$ assumed as for $1^-
  \rightarrow 0^+$; no significant relevance for the \oag\ cross
  section, see Fig.~\ref{fig:DC}.} \\
$3^-$ &  7156 & $4^+$ &  4248 & 1 & $7.9(9)\times 10^{-5}$ &
$5.4\times 10^{-6}$ & 14.7~~~ \\
$5^-$ & 10262 & $4^+$ &  4248 & 1 & $-$ &
$6.0\times 10^{-6}$ & 2.13\footnotemark[1] \\
\hline
\end{tabular}
\end{center}
\end{table}

The above simple two-body model reproduces the partial widths
$\Gamma_\alpha$ of the states under consideration with reasonable
accuracy (see Table \ref{tab:bound}), and the calculated reduced
transition strengths for $E2$ transitions are in good agreement with
the experimental values (see Table \ref{tab:BE}). A prerequisite for
the successful application of a simple two-body model is that the
relevant states in $^{20}$Ne have a dominating $^{16}$O $\otimes$
$\alpha$ cluster structure. Such a structure can be expected from the
double shell closure at $^{16}$O, and it is confirmed by large
spectroscopic factors close to unity which have been extracted from
the analysis of transfer reactions. However, it is difficult to specify
precise values for the spectroscopic factors as can be seen from the
results given in compilations \cite{Til98,Ajz87,Ajz83,Ajz78} and from
recent experiments (see e.g.\ \cite{Mao96,Tan81a,Tan81b,Ana79,Bra79}).

\section{\label{sec:results}Results}
\subsection{\label{sec:sfac}The astrophysical \sfac }
\begin{figure}[hbt]
\includegraphics[ bb = 140 65 490 535, width = 100 mm, clip]{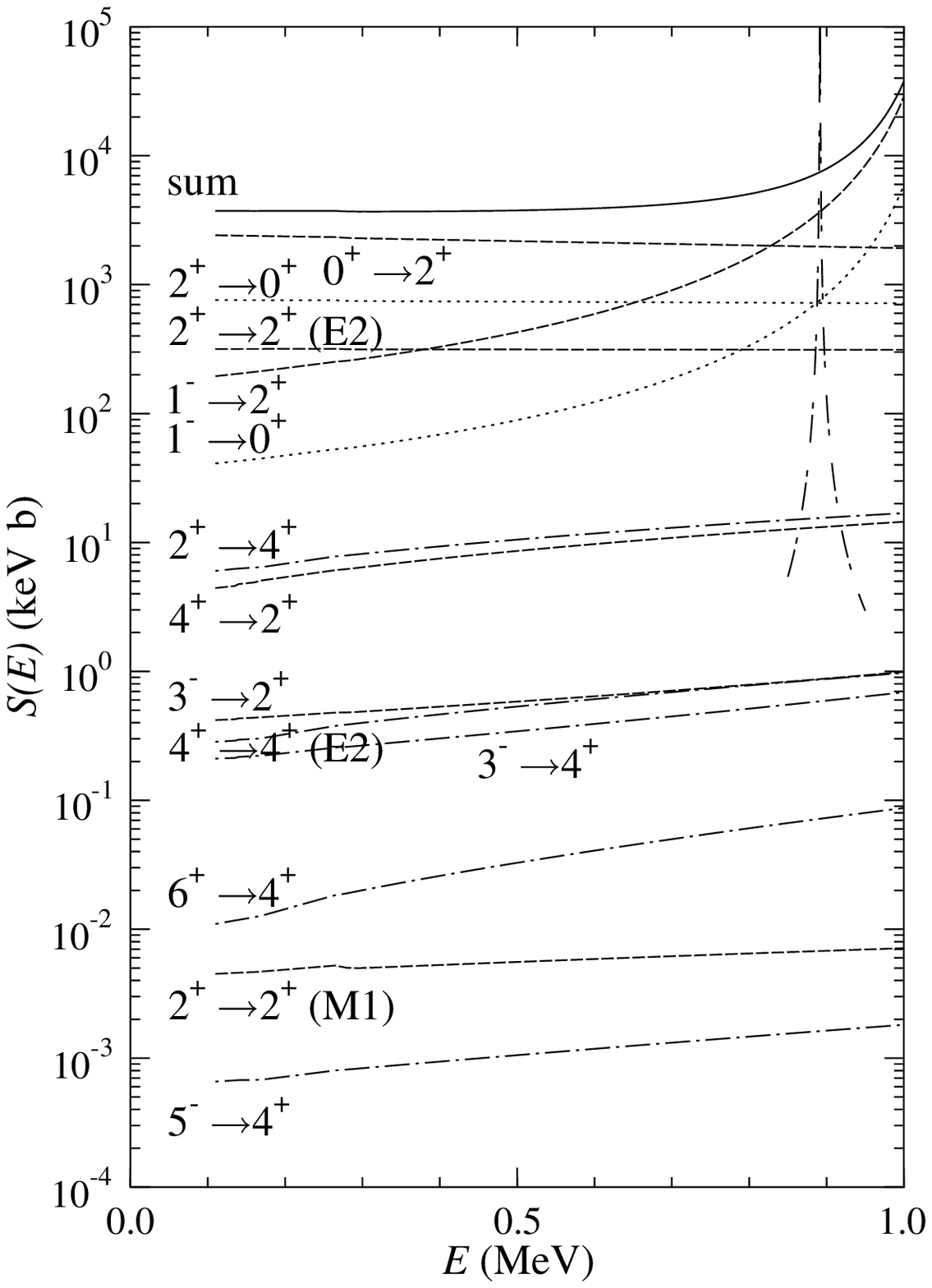}
\caption{
\label{fig:DC} 
\sfac\ of the DC cross section of the \oag\ capture reaction. The sum
of all transitions is shown as full line. Transitions to the $0^+$
ground state ($2^+$ first excited state; $4^+$ second excited state)
in $^{20}$Ne are shown as dotted (dashed; dash-dotted) lines. The $M1$
transition from the incoming $L = 4$ partial wave to the $J^\pi = 4^+$
bound state has a \sfac\ of less than $6 \times 10^{-6}$\,keV\,b (off
scale). The \sfac\ of the narrow $3^-$ resonance at $E = 891$\,keV has
been calculated using the Breit-Wigner formula (dashed line); this
resonance is not included in the model space, see Sect.~\ref{sec:disc}.
}
\end{figure}

\begin{figure}[hbt]
\includegraphics[ bb = 140 65 490 335, width = 100 mm, clip]{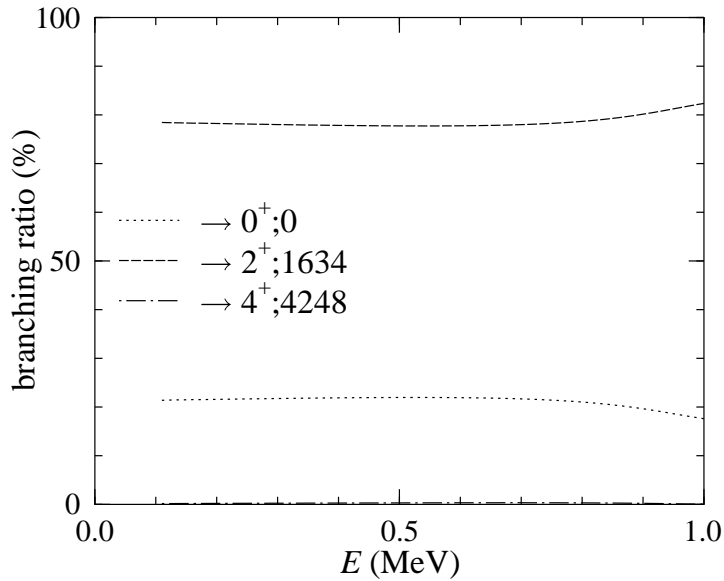}
\caption{
\label{fig:branch} 
Branching ratios of the DC cross section of the \oag\ capture reaction
leading to the $0^+$ ground state (dotted line), $2^+$ state at $E_x =
1634$\,keV (dashed line), and $4^+$ state at $E_x = 4248$\,keV
(dash-dotted line; below 1\,\%, practically not visible).
}
\end{figure}

The calculation of the capture cross section of the \oag\ capture
reaction is now straightforward. The folding potentials (see
Sect.~\ref{sec:fold}) have been adjusted in strength to reproduce the
bound state energies (see Sect.~\ref{sec:bound}). Consequently, the
wave functions of the DC model (see Sect.~\ref{sec:DC}) are
well-defined. The absolute value of the capture cross section is
adjusted using the ratios between the experimental and calculated
$B(E{\cal{L}})$ values (see Eq.~(\ref{eq:ratio}) and
Sect.~\ref{sec:BE}). The results are shown in Figs.~\ref{fig:DC},
\ref{fig:branch}, and \ref{fig:extrapol}.

Fig.~\ref{fig:DC} summarizes the calculations. All possible $E1$,
$E2$, and $M1$ transitions from the incoming partial waves with $0 \le
L \le 6$ to the three bound states with $J^\pi = 0^+$, $2^+$, and
$4^+$ have been taken into account. Below $E = 500$\,keV, about two
third of the total cross section comes from the incoming $L = 0$
partial wave to the first excited $2^+$ state. Further significant
contributions from the incoming $L=2$ partial wave to the ground state
and to the first excited state are of the order of about 20\,\% and
10\,\%. The DC cross section below $E = 500$\,keV is almost entirely
given by $E2$ transitions.

At energies above $E = 500$\,keV the tail of the $1^-$ resonance at $E
= 1058$\,keV becomes more and more important. Although the width of
this resonance is slightly overestimated by the potential model
calculation (see Table \ref{tab:bound}), the adjustment of the
$B(E{\cal{L}})$ values according to Eq.~(\ref{eq:ratio}) assures that
the calculated resonance strength is in agreement with the
experimental value.

Transitions from partial waves with angular momentum $L > 2$ are
practically not relevant for the \oag\ cross section below 1\,MeV. These
transitions are suppressed because of the centrifugal barrier and
because of the reduced transition energy for all transitions to the
second excited state with $J^\pi = 4^+$ at $E_x = 4248$\,keV. The tail
of the $3^-$ resonance at $E = 2426$\,keV is not visible at energies
below $E = 1$\,MeV.

The branching ratios to the ground states with $J^\pi = 0^+$ and the
first (second) excited state with $J^\pi = 2^+$ ($4^+$) at $E_x =
1634$\,keV (4248\,keV) are shown in Fig.~\ref{fig:branch}. The
branching ratios are almost independent of energy below $E = 800$\,keV
with a strong 78\,\% contribution to the $2^+$ state and a 22\,\%
contribution to the $0^+$ ground state. The contribution of the second
excited $4^+$ state remains below 1\,\% in the whole energy range
under consideration.

Experimentally upper limits have been determined for the \sfac\ to the
$0^+$ ground state ($S \rightarrow 0^+ < 920$\,keV\,b) and to the
$2^+$ state ($S \rightarrow 2^+ < 2450$\,keV\,b) \cite{Kn94} from
measurements between resonances at higher energies. The predictions
from the DC model calculation are $S \rightarrow 0^+ = 802$\,keV\,b
and $S \rightarrow 2^+ = 2877$\,keV\,b at $E = 300$\,keV. Because of
the energy dependence of the \sfac , the calculated values do not
contradict the experimental limits of \cite{Kn94}. 
In a similar way as in \cite{Kn94} a total \sfac\ of
$S(300\,{\rm{keV}}) = 700 \pm 
300$\,keV\,b is derived in \cite{Ha87}. However, the
result of \cite{Ha87} has been questioned by \cite{Ba88b}: the
experimental data of \cite{Ha87} at energies around 2\,MeV were
extrapolated to 300\,keV using the probably unrealistic energy
dependence of the \sfac\ from \cite{La83} (see discussion in
\cite{Ba85,Ba88b}); it is stated that this result ``should at least be
multiplied by a factor of two'' \cite{Ba88b}. Additionally, the upper
limit of \cite{Ha87} was derived close to the $0^+$ resonance at
1991\,keV; consequently the derived upper limit depends sensitively on
the chosen
resonance parameters (strength and width) of this resonance \cite{Kn94}.

An estimate of the theoretical uncertainties can be done in the
following way. Obviously, the calculated cross sections, which have
been adjusted to experimental $B(E{\cal{L}})$ values according to
Eq.~(\ref{eq:ratio}), depend on the experimental $B(E{\cal{L}})$
values or $\Gamma_\gamma$ partial widths. Because $\Gamma_\alpha \gg
\Gamma_\gamma$ for all resonances under consideration, the radiation
width $\Gamma_\gamma$ may also be obtained from the resonance strength
$\omega \gamma$ of a resonance in the \oag\ capture reaction. Typical
experimental uncertainties for the resonance strengths are of the
order of about 15\,\% \cite{NACRE}, and results from different
experiments also agree with each other at the 15\,\% level. Combining
these uncertainties, this leads to a total uncertainty of about 20\,\%
for the experimental $B(E{\cal{L}})$ values or radiation widths
$\Gamma_\gamma$. A second source of uncertainties comes from the
adjustment of the potential strength to the binding energies of
(quasi-)bound states (see Table \ref{tab:bound}). The numerical adjustment
itself can be done with negligible uncertainties; however, the model
assumes a pure two-body configuration for all states, and
consequently, the adjustment of the potential may be somewhat
uncertain. But it has ben shown in \cite{Abe93}, that the potential
reproduces experimental phase shifts with reasonable
quality. Additionally, almost the same potential strength parameter
$\lambda$ has been found for all states with positive parity; an
uncertainty of about 2\,\% can be seen from Table \ref{tab:bound}. The
dependence of the \sfac\ or capture cross section on the potential
strength was calculated using a variation of the parameter $\lambda$
of $\pm 2\,\%$. For the dominating transition from the incoming $L =
0$ partial wave to the $2^+$ bound state one obtains a variation of
the \sfac\ of $+14$\,\% ($-15$\,\%) for a 2\,\% enhanced (reduced)
potential strength. Combining all the above uncertainties, a careful
estimate of the total uncertainty of the \sfac\ is about 30\,\%.

\subsection{\label{sec:rate}The astrophysical reaction rate
  $N_A \left\langle \sigma \, v \right\rangle$}
\begin{figure}[hbt]
\includegraphics[ bb = 140 65 490 335, width = 100 mm, clip]{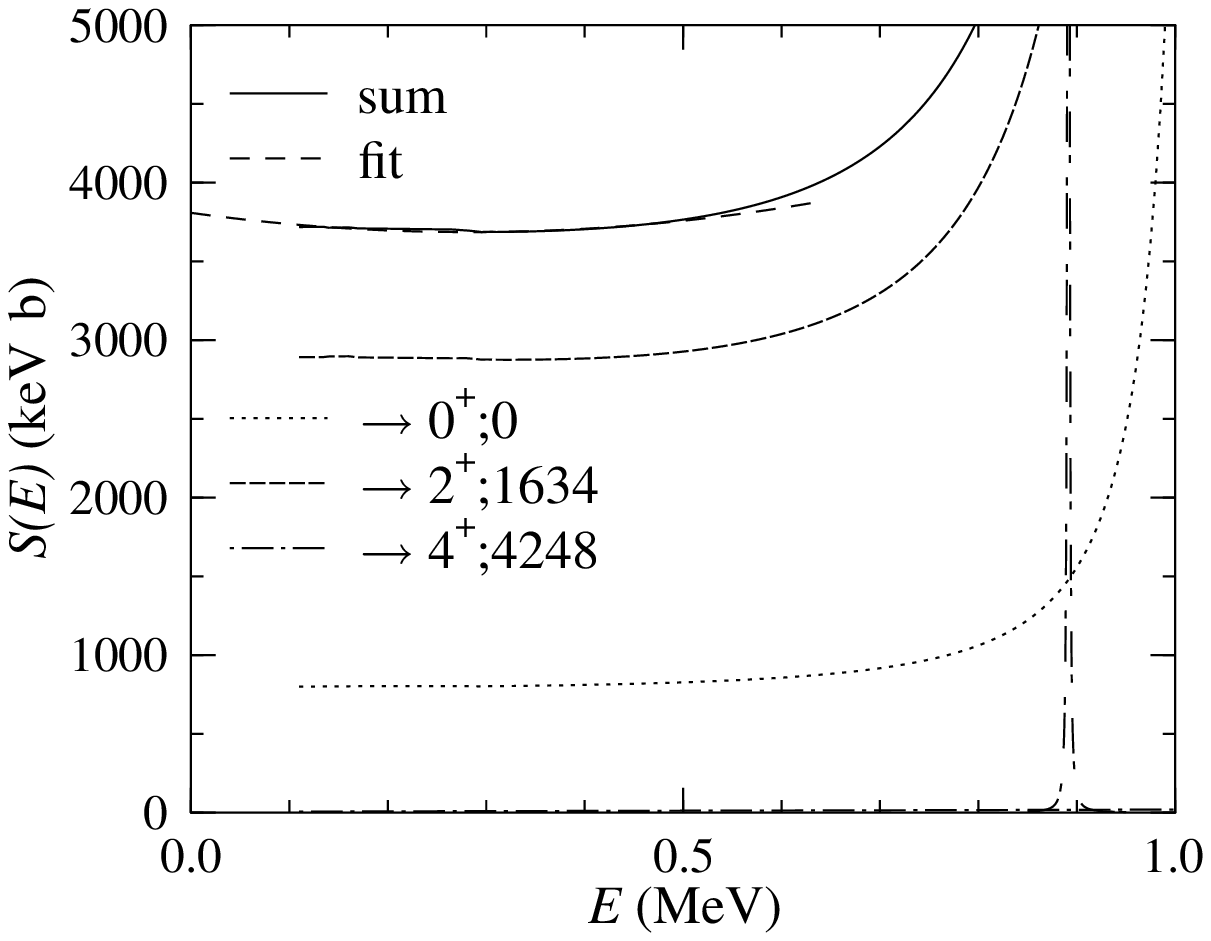}
\caption{
\label{fig:extrapol} 
\sfac\ of the \oag\ reaction in linear scale (same as
Fig.~\ref{fig:DC}). For extrapolation to $E
\rightarrow 0$, the \sfac\ has been fitted using a second-order
polynom $S(E) = 3808\,{\rm{keV\,b}} - 864\,{\rm{keV\,b}} \times
E/{\rm{MeV}} + 1527\,{\rm{keV\,b}} \times (E/{\rm{MeV}})^2$ up to $E =
0.5$\,MeV (long-dashed line). The \sfac\ of the narrow $3^-$ resonance
at $E = 891$\,keV has been calculated using the Breit-Wigner formula
(dashed line); this resonance is not included in the model space, see
Sect.~\ref{sec:disc}. 
}
\end{figure}

For the calculation of the astrophysical reaction rate $N_A
\left\langle \sigma \, v \right\rangle$ in the temperature range where
DC is dominating, one needs the \sfac\ from very low energies to about
500\,keV. A second-order polynom has been fitted to the calculated
\sfac\ which is shown in Fig.~\ref{fig:extrapol} in linear scale:
\begin{equation}
S(E) = 3808\,{\rm{keV\,b}} - 864\,{\rm{keV\,b}} \times
E/{\rm{MeV}} + 1527\,{\rm{keV\,b}} \times (E/{\rm{MeV}})^2
\label{eq:extrapol}
\end{equation}

\begin{figure}[hbt]
\includegraphics[ bb = 140 65 490 250, width = 100 mm, clip]{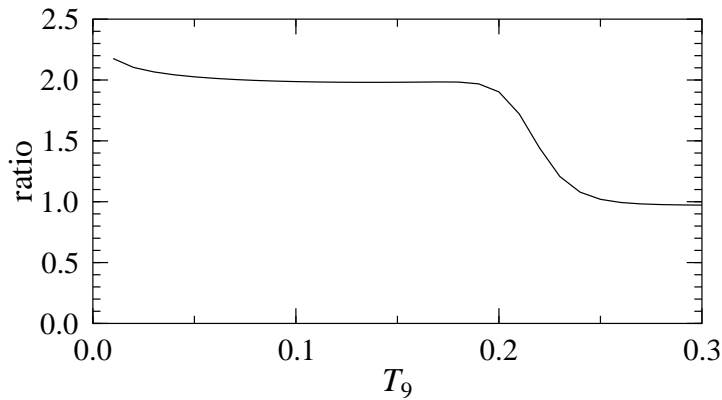}
\caption{
\label{fig:rate} 
Ratio between the reaction rate $N_A \left\langle \sigma \, v
\right\rangle$ calculated from the DC model and folding potentials
(this work) and the NACRE compilation \cite{NACRE}. Because of the
increased DC cross section at low energies, the reaction rate is about
a factor of two larger than the NACRE rate below $T_9 = 0.2$ where DC
is dominating.
}
\end{figure}

Because of the weak energy dependence of the \sfac\ the reaction rate
can be calculated using the approximation
\begin{equation}
N_A \left\langle \sigma \, v \right\rangle =
N_A \left( \frac{2}{\mu} \right)^{1/2} \,
\frac{\Delta}{(kT)^{3/2}} \, S(E_0) \,
\left( 1 + \frac{5}{12\tau} \right) \,
\exp{(-\tau)}
\label{eq:rate}
\end{equation}
with $E_0$ and $\Delta$ as defined in Sect.~\ref{sec:intro} and $\tau
= 3E_0/(kT)$. Compared to \cite{NACRE}, a somewhat larger reaction
rate is obtained because of the larger \sfac\ at the relevant energies
(see Fig.~\ref{fig:rate}). 

At temperatures below $T_9 \approx 0.2$ the reaction rate is almost
entirely given by the DC contribution. Because of the increased \sfac\
at low energies, the reaction rate is enhanced by the same factor. The
almost constant ratio below $T_9 = 0.2$ in Fig.~\ref{fig:rate} arises
from the fact that the energy dependence of the \sfac\ is practically
the same in this work and in \cite{NACRE} which was adopted from
\cite{Du94} (see also Sect.~\ref{sec:disc}). At higher energies the
reaction rate is dominated by the resonant contributions, and hence
the increased DC cross section has no significant influence on the
reaction rate.

No distinction is made here between the reaction rates $N_A \left\langle
\sigma \, v \right\rangle^\star$ under stellar conditions and $N_A
\left\langle \sigma \, v \right\rangle_{\rm{lab}}$ in the
laboratory. Both rates are practically identical up to temperatures of
about $T_9 = 3$ \cite{NACRE}. There is not significant thermal
population of excited states in $^{16}$O which are located at
excitation energies above $E_x = 6$\,MeV. The stellar reaction rate
$N_A \left\langle \sigma \, v \right\rangle^\star$ can directly be derived
from experimental data.

\section{\label{sec:disc}Discussion}
Because of the properties of the doubly-magic $^4$He and $^{16}$O
nuclei a simple two-body model should be able to describe scattering
phase shifts, bound state properties of $^{20}$Ne, and the capture
cross section of the \oag\ reaction with reasonable accuracy
\cite{De83,La83,Ba85}. The combination of the model with systematic
folding potentials ensures a good description of elastic scattering
cross sections and phase shifts at low energies.

The strength parameter $\lambda$ is adjusted to the binding energies
of low-lying states in $^{20}$Ne; this guarantees the correct
asymptotic behavior of the bound state wave functions at large
distances which is important for the calculation of the \oag\ capture
cross section. The correct behavior of the wave functions at small
distances, i.e.\ the number of nodes of the wave function and their
positions, is obtained 
from Eq.~(\ref{eq:Q}) which avoids Pauli-forbidden states. The precise
influence of antisymmetrization of the 20-nucleon wave function has
been discussed in detail in \cite{Ba85}; it has been pointed out ``the
smallness of the residual antisymmetrization terms'' because of the
shell closures in $^{4}$He and $^{16}$O.

The potential in the chosen model automatically contains eigenstates
at energies below and above the $^{16}$O-$\alpha$ threshold. The
quasi-bound states above the threshold appear as resonances in the
\oaa\ scattering phase shifts and in the \oag\ capture cross
section. As already pointed out in \cite{Ba85}, the term ``direct
capture'' may be misleading in this case. Further details on such
so-called potential resonances are given in
\cite{Mohr94,Mohr99,Mohr03}. However, the model contains only a
limited number of eigenstates. It is not possible to include all
resonances of the \oag\ reaction in this simple two-body model. E.g.,
the states with dominating $^{16}$O($3^-$) $\otimes$ $\alpha$
structure at $E_x = 4967$\,keV ($E = 237$\,keV; no resonance because
of the unnatural $2^-$ parity) and $5621$\,keV ($E = 891$\,keV) are not
included in the model; the inclusion of such states requires e.g.\ a
multicluster model \cite{Du94}. The influence of the $3^-$ resonance
at 891\,keV at low energies is negligible as can be seen from
Figs.~\ref{fig:DC} and \ref{fig:extrapol}. The \sfac\ of this
resonance has been calculated from the Breit-Wigner formula using the
known widths and strength \cite{Til98,NACRE}.

The calculation of the overlap integrals in Eq.~(\ref{eq:DC}) depends
sensitively on the shape of the wave functions especially for wave
functions with several nodes. Minor shifts of the nodes may reduce or
enhance the overlap integral significantly because of cancellation
effects between positive and negative regions of the integrand in
Eq.~(\ref{eq:DC}) (see Figs.~2-5 of Ref.~\cite{Ba85}). This leads to a
relatively sensitive dependence of the \oag\ cross section on the
chosen potential, i.e.\ the potential strength parameter $\lambda$. A
variation of $\lambda$ by its uncertainty of $\pm 2\,\%$ (see Table
\ref{tab:bound}) changes the capture cross section already by about
15\,\% which has to be considered as the intrinsic model
uncertainty. Further uncertainties as discussed in
Sect.~\ref{sec:sfac} lead to a total uncertainty of the calculated
\sfac\ of about 30\,\%.

The adjustment of the DC cross sections of the \oag\ reaction using
the ratios between experimental and theoretical $B(E{\cal{L}})$
according to Eq.~(\ref{eq:ratio}) gives the chance to include E1
transitions in the DC model. As stated earlier (see
Eqs.~(\ref{eq:isoE1}), (\ref{eq:betaL}), and Sect.~\ref{sec:model}),
E1 transitions are suppressed in the DC model. A detailed microscopic
investigation of E1 transitions in the \oag\ reaction was able to
reproduce the approximate strength of the transitions, but showed
problems with the branching ratios \cite{De86}. Therefore the usage of
Eq.~(\ref{eq:ratio}) seems to be the best way to predict the E1 part
of the \oag\ cross section at low energies: the energy dependence of
the cross section is provided by the model, and the absolute strength
and the branching ratio are obtained from the adjustment to
experimental data. The total contribution of E1 transitions to the
\oag\ cross section remains small as long as energies significantly
below the $1^-$ resonance at 1058\,keV are considered.

The final result for the total \sfac\ at 300\,keV is
$S(300\,{\rm{keV}}) = 3.7 \pm 1.2$\,MeV\,b which is slightly higher
than previous calculations \cite{Du94,Ba88b,De86,Ba85,De83,La83} but
overlaps with the range of the results of
\cite{Du94,Ba88b,De86,Ba85,De83} within the estimated
uncertainties. The result of \cite{La83} is significantly lower but
has been questioned in \cite{Ba85}. The calculated slope $a$ of the
\sfac\ at low energies, usually defined as
\begin{equation}
a = \frac{1}{S(0)} \, \left(\frac{dS}{dE}\right)_{E=0}
\label{eq:slope}
\end{equation}
is in good agreement with previous calculations: $a =
-0.23$\,MeV$^{-1}$ from this work for the total \sfac\ is almost
identical to $a = -0.21$\,MeV$^{-1}$ for the dominating $L=0$
$\rightarrow$ $2^+$ transition in \cite{Ba85}. The \sfac\ decreases
slightly with increasing energy in all calculations
\cite{Du94,Ba88b,De86,Ba85,De83}. The only exception is the result
of \cite{La83} who finds an increasing \sfac\ with increasing energy
with values of $a \approx +0.6$\,MeV$^{-1}$. A possible shortcoming of
\cite{La83} is discussed in \cite{Ba85}.

Branching ratios $b$, here often defined as
\begin{equation}
b = \frac{S_0+S_2}{S_0}
\label{eq:Sbranch}
\end{equation}
can be estimated from Fig.~5 of \cite{Du94} to be of the order of $3
\alt b \alt 3.5$, and $b \approx 4.3$ is given in \cite{Ba88b}. In
this work $b = 4.5$ is found. Again this calculation agrees nicely
with the microscopic calcaulations by \cite{Du94,Ba88b}.

Any comparison with experimental data is difficult. No experimental
data are available at energies below 1\,MeV, and the extraction of the
DC contribution from experimental data at higher energies is hampered
by the dominating resonant contributions. As pointed out in
\cite{Kn94}, the extraction of the DC cross section depends sensitively
on the chosen resonance parameters, because the DC cross section is usually
determined as the difference between the measured cross section and
the calculated resonant cross section in the tail of a resonance with
Breit-Wigner shape. As pointed out already in Sect.~\ref{sec:sfac}, the new
calculation does not contradict the upper limits of \cite{Kn94}.

Finally, the astrophysical reaction rate $N_A \left\langle \sigma \, v
\right\rangle$ is about a factor of two higher than in the NACRE
compilation \cite{NACRE} below $T_9 = 0.2$ because the \sfac\ at low
energies is about 3.7\,MeV\,b in this work compared to the adopted
value of 2\,MeV\,b in \cite{NACRE}. A detailed analysis of the
astrophysical consequences is beyond the scope of this paper; in
general, the \oag\ reaction rate is still low enough to block the
reaction chain 3\,$\alpha$ $\rightarrow$ $^{12}$C\rag \oag . Using the
NACRE rate for the $^{12}$C\rag $^{16}$O reaction and the new rate for
the \oag\ reaction, one finds at $T_9 = 0.2$ a ratio of about $5
\times 10^{-5}$ between the \oag\ and the $^{12}$C\rag $^{16}$O
rates. For comparison, abundances calculated from the NACRE rates can
be found in \cite{Arn99}.

\section{\label{sec:inv}The inverse $^{20}$N{\lowercase{e}}\rga
  $^{16}$O Reaction} 
Reverse reaction rates $\lambda^\star$ under stellar conditions are
usually derived from the detailed balance theorem \cite{NACRE,Fo67}
which reads for the case of the \oag\ and \nega\ reactions
\begin{equation}
\frac{\lambda^\star}{\left\langle \sigma v \right\rangle} 
= \left(\frac{\mu \, kT}{2\pi
  \hbar^2}\right)^{3/2} \, \exp{\left(-\frac{Q}{kT}\right)} 
\label{eq:rate_ratio}
\end{equation}
because the partition functions do not deviate significantly from
unity for the nuclei $^4$He, $^{16}$O, and $^{20}$Ne up to
temperatures of $T_9 \approx 3$ \cite{NACRE} because of the large
excitation energies of the first excited states ($^{16}$O: $E_x =
6049$\,keV, $0^+$; $^{20}$Ne: $E_x = 1634$\,keV, $2^+$). The reaction
$Q$-value is $Q = +4730$\,keV. It has been
shown in \cite{Mohr05} that the detailed balance theorem remains valid
for resonant transitions although the thermal occupation probability
is extremely small for the first excited state in $^{20}$Ne and
negligible for $^{16}$O for all astrophysically relevant temperatures.

The same arguments as given in \cite{Mohr05} hold for the non-resonant
case at temperatures below $T_9 = 0.2$ (although the
photodisintegration rate is practically negligible at such low
temperatures). The reaction rate $\lambda_i$ of the \nega\
photodisintegration reaction for a given state $i$ in $^{20}$Ne is
given by
\begin{eqnarray}
\lambda_i & = & c \int_{Q-E_{x,i}}^{\infty} n_\gamma(E_\gamma,T) \,
\sigma_i^{(\gamma,\alpha)}(E_\gamma) \, dE_\gamma \nonumber \\
& \approx & \frac{c}{\pi^2} \frac{1}{(\hbar c)^3} \frac{\mu c^2}{(2I_i+1)} 
\exp{\left(\frac{+E_{x,i}-Q}{kT}\right)} \int_0^\infty
\exp{\left(-\frac{E_\alpha}{kT}-2\pi\eta\right)} \, S_i(E_\alpha) \,
dE_\alpha 
\label{eq:rate_invers}
\end{eqnarray}
with the thermal photon density 
\begin{equation}
n_\gamma \, (E_\gamma,T) \, dE_\gamma = 
	\frac{1}{\pi^2} \, \frac{1}{(\hbar c)^3} \, 
	\frac{E_\gamma^2}{\exp{(E_\gamma/kT)} - 1} \, dE_\gamma
\label{eq:planck}
\end{equation}
(see e.g.\ \cite{Mohr00}), the excitation energy $E_{x,i}$ and the
spin $I_i$ of the $i$-th state, the Sommerfeld parameter $\eta$, and
the astrophysical \sfac\ $S_i$ for the transition to the $i$-th state
in $^{20}$Ne. The derivation of Eq.~(\ref{eq:rate_invers}) uses only
the time reversal symmetry for the cross sections of transitions
between individual states \cite{Fo67,Bl52}.
Note that the energies are are related by
\begin{equation}
E_\gamma = Q + E_\alpha - E_{x,i}
\label{eq:energies}
\end{equation}
with $E_\alpha$ in the center-of-mass system.
This means that the required photon energy $E_\gamma$ for the
photodisintegration of $^{20}$Ne is reduced by the excitation energy
$E_{x,i}$ of the $i$-th state in $^{20}$Ne. Consequently, the reaction
rate $\lambda_i$ of the $i$-th state is enhanced by a factor
$\exp{\left(+\frac{E_{x,i}}{kT}\right)}$ because of the enhanced
photon density at lower photon energies.

After summing all the rates according Eq.~(\ref{eq:rate_invers}) with
their proper thermal weights proportional to the Boltzmann factor
$\exp{\left(-\frac{E_{x,i}}{kT}\right)}$ 
under stellar conditions one obtains a stellar reaction rate 
\begin{equation}
\lambda^\star
\approx \frac{c}{\pi^2} \, \frac{1}{(\hbar c)^3} \, {\mu c^2} \,
\exp{\left(-\frac{Q}{kT}\right)} \int_0^\infty
\exp{\left(-\frac{E_\alpha}{kT}-2\pi\eta\right)} \,
S_{\rm{tot}}(E_\alpha) \, dE_\alpha 
\label{eq:rate_invers_stellar}
\end{equation}
with the total astrophysical \sfac\ $S_{\rm{tot}} = \sum_i S_i$ for the \oag\
capture reaction. The comparison of Eq.~(\ref{eq:rate_invers_stellar})
with the usual definition of the
reaction rate of the capture reaction leads directly to the detailed
balance result in Eq.~(\ref{eq:rate_ratio}). By the way, the analysis
of the integrand in Eq.~(\ref{eq:rate_invers_stellar}) together with
Eq.~(\ref{eq:energies}) shows that the most effective energy for the
inverse photodisintegration reaction is located at
$E_x(^{20}{\rm{Ne}}) = Q + E_0$ where $E_0$ is the most effective
energy of the capture reaction as defined in
Sect.~\ref{sec:intro}. Further details on the Gamow window for \rga\
reactions see \cite{Mohr03b}.

Although their thermal occupation probabilities are extremely low
because of the Boltzmann factor, excited states play a significant
role in the photodisintegration process because the Boltzmann factor
is exactly balanced by an enhancement factor because of the lower
required photon energy according to Eqs.~(\ref{eq:planck}) and
(\ref{eq:energies}). In the case of the \nega\ photodisintegration
reaction, the first excited state in $^{20}$Ne contributes with 78\,\%
to the photodisintegration rate $\lambda^\star$ because of the 78\,\%
branching to this state in the \oag\ capture reaction although the
thermal occupation probability e.g.\ at $T_9 = 0.2$ is below
$10^{-40}$\,! But there is no significant contribution of the first
excited state in $^{16}$O neither to the \oag\ capture reaction nor to
the \nega\ photodisintegration reaction because there is no
compensation in the cross section or reaction rate for the suppression
factor from the Boltzmann statistics.

Experimentally, photodisintegration rates in the laboratory can be
measured using quasi-thermal photon spectra
\cite{Mohr00,Uts05}. Contrary to the capture rate (see
Sect.~\ref{sec:rate}), the photodisintegration rates $\lambda^\star$
under stellar conditions and $\lambda_{\rm{lab}}$ in the laboratory
are not identical: $\lambda^\star \ne \lambda_{\rm{lab}}$\,!
The laboratory experiment provides
$\lambda_{\rm{lab}}$ with the target $^{20}$Ne in its ground state
which is given by Eq.~(\ref{eq:rate_invers}) for $i = {\rm{g.s.}}$;
compared to Eq.~(\ref{eq:rate_invers_stellar}) one finds a ratio of
\begin{equation}
\frac{\lambda_{\rm{lab}}(T)}{\lambda^\star(T)} \approx
\frac{S_{\rm{g.s.}}(E_0)}{S_{\rm{tot}}(E_0)} \approx 0.22
\label{eq:ratio_invers}
\end{equation}
around $T_9 = 0.2$ for the \nega\ reaction. $E_0$ is the most
effective energy of the capture reaction at temperature $T$, see
Sect.~\ref{sec:intro}. Provided that the
branching ratio $S_{\rm{g.s.}}/S_{\rm{tot}}$ is experimentally known, the
stellar rate $\lambda^\star$ can easily be derived from the
experimental result $\lambda_{\rm{lab}}$. Otherwise, theoretical
models have to be used to determine $\lambda^\star$ from
$\lambda_{\rm{lab}}$. In any case, experimental data may provide
helpful information to restrict theoretical calculations of the
laboratory cross section or rate \cite{Rau04}. Further information on
photodisintegration experiments for the \nega\ reaction is given in
\cite{Mohr05}.

\section{\label{sec:summconc}Summary and Conclusions}
The cross section of the \oag\ capture reaction has been calculated
using a simple two-body DC model together with systematic folding
potentials. The adjustment of the potential strength and the reduced
transition probabilities to the relevant experimental quantitites
enables the calculation of all relevant E1, E2, and M1 transitions
between incoming partial waves with $0 \le L \le 6$ and all bound
states ($J^\pi = 0^+$, $E_x = 0$; $2^+$, 1634\,keV; and $4^+$,
4248\,keV) in $^{20}$Ne without further free parameters.

The astrophysical \sfac\ at energies below 1\,MeV is somewhat higher
than in previous calculations \cite{Du94,Ba88b,De86,Ba85,De83},
whereas the energy dependence and the branching ratio to the final
states is in good agreement. Consequently, the astrophysical reaction
rate is enhanced compared to a recent compilation \cite{NACRE} at
temperatures below $T_9 = 0.2$ where the direct capture contribution
is dominating.

The astrophysical reaction rate of the inverse \nega\
photodisintegration reaction has a significant contribution of the
thermally excited first excited state in $^{20}$Ne although the
thermal occupation probability seems to be negligible. The widely used
detailed balance theorem remains valid also for the case of high-lying
first excited states.

\begin{acknowledgments}
I thank C.\ Angulo, P.\ Descouvemont, Zs.\ F\"ul\"op, G.\ Staudt, and
H.\ Utsunomiya for encouraging discussions.
\end{acknowledgments}

\end{document}